\documentclass[
    aps,
    prl,
    twocolumn,
    superscriptaddress,
    showpacs,
    nofootinbib,
    floatfix
]{revtex4-2}

\usepackage{graphicx}
\usepackage{amsmath}
\usepackage{amssymb}
\usepackage{amsfonts}
\usepackage{braket}
\usepackage{mathtools}
\usepackage{bm}
\usepackage[colorlinks=true,allcolors=blue]{hyperref}
\usepackage{float}
\usepackage[justification=justified]{caption}
\usepackage{subcaption}

\begin{document}

\title{{\em Ab initio} {\boldmath$\alpha$}--{\boldmath$\alpha$} scattering with high-fidelity chiral interactions}

\author{Avik Sarkar}
\email{sarkar.avik.d6@tohoku.ac.jp}
\affiliation{Department of Physics, Tohoku University, Aoba 6-3, Sendai, 980-8578, Miyagi, Japan}
\affiliation{Institute for Advanced Simulation (IAS-4),
 Forschungszentrum J\"ulich, D-52425 J\"ulich, Germany}
\affiliation{Helmholtz Institut f\"ur Strahlen- und Kernphysik and Bethe Center for Theoretical Physics,\\ Universit\"at Bonn, D-53115 Bonn, Germany}

\author{Serdar Elhatisari}
\email{selhatisari@gmail.com}
\affiliation{Faculty of Natural Sciences and Engineering, Gaziantep Islam Science and Technology University, Gaziantep 27010, Turkey}

\author{Timo A. L\"ahde}
\email{t.laehde@fz-juelich.de}
\affiliation{Institute for Advanced Simulation (IAS-4),
Forschungszentrum J\"ulich, D-52425 J\"ulich, Germany}

\author{Ulf-G. Mei{\ss}ner}
\email{meissner@hiskp.uni-bonn.de}
\affiliation{Helmholtz Institut f\"ur Strahlen- und Kernphysik and Bethe Center for Theoretical Physics,\\ Universit\"at Bonn, D-53115 Bonn, Germany}
\affiliation{Institute for Advanced Simulation (IAS-4),
 Forschungszentrum J\"ulich, D-52425 J\"ulich, Germany}
\affiliation{Peng Huanwu Collaborative Center for Research and Education, International Institute for Interdisciplinary and Frontiers, Beihang University, Beijing 100191, China}

\date{\today}

\begin{abstract}
Low-energy $\alpha$--$\alpha$ scattering underlies stellar helium burning and sharply tests nuclear forces in the reaction regime. We present its first calculation using the high-fidelity N$^{3}$LO chiral NLEFT interaction, incorporated through wave function matching, on a fine 
lattice, using the adiabatic projection method. On the fine lattice, the two-cluster norm matrix becomes severely ill-conditioned, and its direct inversion is unstable. We address this with Tikhonov regularization, extrapolating the regulator to zero, and confirm the result with an independent truncated singular-value decomposition.  The $S$- and $D$-wave phase shifts agree with empirical analyses, extending the validation of this interaction from bound states and charge radii to scattering and providing a practical route to {\em ab initio} nuclear reactions on fine lattices.

\end{abstract}

\maketitle

\section{Introduction}

The elastic scattering of two $\alpha$ particles is one of the most fundamental processes in low-energy nuclear physics. The $^{8}$Be system it probes is unbound, decaying back into two $\alpha$ particles, yet its narrow $0^{+}$ ground-state resonance just above the $\alpha$--$\alpha$ threshold and its broad $2^{+}$ first excited state are the gateway to almost all He burning in stars. The triple-$\alpha$ process proceeds through the unstable $^{8}$Be ground state followed by radiative capture of a third $\alpha$ particle into the Hoyle state of $^{12}$C~\cite{Hoyle1954, Burbidge1957, Fowler1984, FreerFynbo2014}, and subsequent $\alpha$ captures generate the nuclear ``ash'' of O, Ne, and heavier elements that shape stellar evolution and type-Ia supernovae~\cite{Freer2018}. Because the $^{8}$Be resonances are imprinted directly on the $\alpha$--$\alpha$ phase shifts, elastic scattering provides a clean, parameter-free benchmark for any microscopic theory of $\alpha$-cluster dynamics~\cite{Afzal1969}.

Empirical $\alpha$--$\alpha$ phase shifts have long been available from phenomenological analyses~\cite{Afzal1969}, but a predictive microscopic description remains challenging, for two reasons. First, the long-lived $0^{+}$ resonance and the strong Pauli antisymmetrization between the two clusters make $\alpha$--$\alpha$ a genuinely eight-body problem rather than a structureless two-body one. Second, the very feature that makes $\alpha$ clusters interesting---the delicate competition between the formation of compact clusters and their mutual interaction---is extremely sensitive to the details of the underlying nucleon--nucleon and many-nucleon forces, in particular to their off-shell behavior and locality~\cite{Elhatisari2016QPT}. A realistic calculation of $\alpha$--$\alpha$ scattering is therefore a far more demanding test of a nuclear interaction than reproducing a binding energy or a charge radius.

\textit{Ab initio} nuclear theory has made striking progress on light-nucleus reactions over the past two decades, but almost always with a light probe on a light target. Exact few-body methods, such as the Faddeev-Yakubovsky equations and the hyperspherical harmonics expansion, treat the scattering problem essentially without approximation, but are limited to four or five nucleons~\cite{Lazauskas2018, LazauskasCarbonell2020, Kievsky2008, Marcucci2020}. Coordinate-space quantum Monte Carlo describes bound states and narrow resonances up to $A\simeq 12$~\cite{Carlson2015}, has been extended to $n$--$\alpha$ scattering~\cite{Nollett2007}, and reproduces the $A=8$ spectrum, including the $^{8}$Be rotational band, as resonances~\cite{Wiringa2000}. The no-core shell model with continuum couples many-body states to binary-cluster channels~\cite{QuaglioniNavratil2008, Baroni2013, Navratil2016, Kravvaris2023}, yielding a wide range of reactions and astrophysical $S$-factors, and has recently reached the $\alpha$--$\alpha$ system and the structure of $^{8}$Be~\cite{Kravvaris2024}. Coupled-cluster theory with continuum degrees of freedom reaches elastic nucleon scattering off medium-mass targets~\cite{Hagen2012}. In all of these approaches, however, the cost of antisymmetrizing and propagating two \emph{heavy, composite} clusters in the asymptotic region grows steeply, so that $\alpha$--$\alpha$ scattering sits at the frontier of what is feasible, and the heavier $\alpha$-cluster reactions of nucleosynthesis remain largely beyond reach.

Nuclear lattice effective field theory (NLEFT) occupies a complementary niche. Its mild computational scaling, combined with the adiabatic projection method (APM)~\cite{Pine2013, Elhatisari2014FD}, formulated directly in terms of composite two-cluster states, enabled the first \textit{ab initio} study of $\alpha$--$\alpha$ scattering~\cite{Elhatisari2015AA} and defines a route that scales favorably toward the heavier $\alpha$-cluster reactions, such as triple-$\alpha$ and radiative $\alpha$ capture, that forge carbon and oxygen in evolved stars. The most recent APM study essentially reproduced the empirical $\alpha$--$\alpha$ $S$-wave phase shift~\cite{Elhatisari2022Multiverse}, though a residual discrepancy persisted in the $D$-wave. As APM itself has since been substantially improved, this tension points to the limitations of the earlier chiral interactions and motivates revisiting $\alpha$--$\alpha$ scattering with a more accurate, modern NLEFT force.

Wave function matching (WFM)~\cite{Elhatisari2024WFM} has since made it possible to treat a realistic chiral interaction at next-to-next-to-next-to-leading order (N$^{3}$LO) with high fidelity on a finer lattice ($a=1.32~\mathrm{fm}$) that strongly reduces rotational-symmetry-breaking artifacts~\cite{Lu:2015gfa}. This interaction reproduces a broad set of observables---nucleon-nucleon phase shifts, the binding energies and charge radii of nuclei up to $A\leq 58$, and the neutron-matter equation of state and saturation properties~\cite{Elhatisari2024WFM}---and has since passed further tests, including the spectroscopy of the Be isotopes and the ground-state parity inversion of $^{11}$Be~\cite{Shen2025Be} and the properties of hot neutron matter~\cite{Ma:2023ahg}. All of these, however, probe bound-state or infinite-matter observables. Scattering is a complementary and more stringent probe, sensitive to the interaction away from the energetically favored bound configurations and to the off-shell and clustering properties that bound states constrain only weakly~\cite{Elhatisari2016QPT}. No such test of this interaction has yet been carried out for a composite-cluster system.

In this Letter, we provide that test by computing the $S$- and $D$-wave $\alpha$--$\alpha$ phase shifts with the chiral interaction at N$^{3}$LO on a fine lattice. A companion finite-volume study has examined the simpler $n$--$\alpha$ system with the same Hamiltonian~\cite{Elhatisari_2025}, extracting phase shifts with the L\"uscher formalism. Here, we use APM to reduce the eight-nucleon problem to an effective two-cluster system and extract the phase shifts with the spherical-wall method~\cite{Carlson:1984zz, Borasoy2007}. The chief practical obstacle is the ill-conditioning of the APM norm matrix. On the fine lattice the two-cluster basis is finely resolved and the norm matrix spans a wide range of scales, so it becomes severely ill-conditioned. Its direct inversion is then unstable, and in the presence of statistical Monte Carlo (MC) noise it returns phase shifts with uncertainties of order the signal itself. We address this by treating the inversion as an ill-posed problem and regularizing the norm matrix with two independent schemes. Their agreement indicates that the result is controlled by the well-resolved subspace and is insensitive to the noisy low-norm directions. This removes a practical obstacle to applying the APM on fine lattices and, with it, to \textit{ab initio} composite-cluster reactions more generally.

\section{Theoretical framework}
\emph{Nuclear lattice effective field theory.}---NLEFT is by now a mature framework that combines Euclidean-time projection, chiral effective field theory (EFT), and lattice Monte Carlo methods to solve the nuclear many-body problem~\cite{Lee2009, Laehde2019, Lee2025}. As in lattice QCD, it is formulated as a path integral over discretized space and Euclidean time, with the nucleon-nucleon interaction organized as a systematic expansion in powers of momenta and pion masses. Trial wave functions are evolved in Euclidean time to access the spectra, structure, and transitions of low- and medium-mass nuclei, and the mild (polynomial) scaling of the cost with nucleon number $A$ makes phenomena such as $\alpha$ clustering accessible. Prominent successes include the structure of the Hoyle state in $^{12}$C~\cite{Epelbaum2011Hoyle, Epelbaum2012Hoyle}, clustering near a quantum phase transition~\cite{Elhatisari2016QPT}, and the first \textit{ab initio} $\alpha$--$\alpha$ scattering calculation~\cite{Elhatisari2015AA}.

\emph{Wave function matching.}---The interaction used here is implemented through wave function matching~\cite{Elhatisari2024WFM}, which circumvents the slow convergence of perturbation theory and the MC sign problem. WFM applies a short-range unitary transformation that maps a realistic but computationally hard chiral Hamiltonian onto an easily computable one, while leaving the long-range physics untouched. The realistic interaction can then be evaluated to high chiral order in perturbation theory without severe sign oscillations. In practice, we use a high-fidelity chiral interaction at N$^{3}$LO on a fine lattice ($a=1.32~\mathrm{fm}$), built on chiral forces validated against nucleon--nucleon scattering data~\cite{Li2018np, Alarcon2017, Lu2019Essential} (see Sec.~S1 of the Supplemental Material~\cite{SM} for the full lattice action and conventions).

\emph{Adiabatic projection method.}---The APM~\cite{Pine2013, Elhatisari2014FD, Elhatisari2015AA} reduces the underlying multi-nucleon dynamics to an effective two-cluster Hamiltonian. We construct initial trial states $|\vec{R}\,\rangle$, each consisting of two $\alpha$ clusters whose centers are separated by the lattice displacement $\vec{R}$, and project them in Euclidean time,
\begin{equation}
  |\vec{R}\,\rangle_{t} \;=\; e^{-Ht}\,|\vec{R}\,\rangle ,
\end{equation}
so that the ``dressed'' states acquire the full two-cluster correlations, such as deformation and polarization, induced by the microscopic interaction. From these, we form the norm and Hamiltonian matrices,
\begin{align}
  [N(t)]_{\vec{R}\vec{R}'} &= {}_{t\,}\!\langle \vec{R}\,|\vec{R}'\,\rangle_{t},\\
  [H(t)]_{\vec{R}\vec{R}'} &= {}_{t\,}\!\langle \vec{R}\,|\,H\,|\vec{R}'\,\rangle_{t},
\end{align}
and obtain the effective two-cluster adiabatic Hamiltonian by a symmetric orthogonalization of the non-orthogonal dressed basis,
\begin{equation}
  H^{a}(t) \;=\; N^{-1/2}(t)\,H(t)\,N^{-1/2}(t).
  \label{eq:Ha}
\end{equation}
In practice, the displacement states are projected onto definite orbital angular momentum and grouped into radial bins, which organizes the basis into partial-wave channels and keeps the dimensionality minimal~\cite{Elhatisari2016APM}.

Extracting low-energy phase shifts requires a lattice large enough to contain the asymptotic region in which the two clusters are well separated, yet a fully interacting eight-nucleon calculation in such a volume is prohibitively expensive. We therefore build $H^{a}$ in two steps~\cite{Elhatisari2016APM, Elhatisari_2025}. The interacting Hamiltonian of Eq.~\eqref{eq:Ha} is computed on a small lattice, only somewhat larger than the inter-cluster interaction range, retaining the full dynamics of all eight nucleons. Separately, the two $\alpha$ particles are propagated as isolated clusters---intra-cluster forces retained, inter-cluster interaction switched off---on a much larger lattice, which is inexpensive because the clusters never overlap. Beyond the interaction range, the interacting adiabatic Hamiltonian reduces to this free two-cluster one, so the two are matched in the asymptotic region, with the small-box result kept at short cluster separations and the large-box free result at large separations. The merged, large-volume adiabatic Hamiltonian then resolves the low energies needed for the phase shifts without ever requiring an interacting calculation in the large box (see Sec.~S2~\cite{SM}).

After projection onto good angular momentum quantum numbers and radial binning, the phase shifts are computed from $H^{a}$ with the spherical-wall method~\cite{Borasoy2007, Lu2016, Elhatisari2016APM}. Since we include the Coulomb interaction, asymptotic Coulomb wave functions are used in the spherical-wall analysis. The fully interacting calculations are carried out on a (small) cubic box of side $L'=10$ at spacing $a=1.32~\mathrm{fm}$ (momentum cutoff $470~\mathrm{MeV}$), which we merge with the free two-cluster Hamiltonian on a (large) box of side $L=72$ (see also Sec.~S2~\cite{SM}). The two-cluster basis uses $11$ radial bins of width $0.5~\mathrm{fm}$, and statistical MC errors are estimated by jackknife resampling.

\section{Regularization of the norm matrix}
\emph{The conditioning problem.}---Constructing $H^{a}$ requires the inverse square root of $N$. The 2015 $\alpha$--$\alpha$ study~\cite{Elhatisari2015AA} used a coarser lattice ($a=1.97~\mathrm{fm}$) that acts as a strong momentum cutoff. It smooths the stochastic signal and keeps $N$ well conditioned, but at the price of short-distance resolution and chiral order. On the present fine lattice, the spectrum of $N$ spans many orders of magnitude, and its smallest eigenvalues lie at the level of the statistical noise. The construction of $H^{a}$ amplifies these low-norm directions, and a direct inversion yields phase shifts with uncertainties of $\mathcal{O}(10^{2})$ degrees---comparable to the entire physical range---even though the high-norm subspace that carries the physics is accurately sampled and the APM is exact in principle. Resolving the smallest eigenvalues by brute force would require roughly two orders of magnitude more MC statistics than is available. We instead treat the inversion as an ill-posed problem and regularize it with two independent schemes, and their agreement confirms the result.

\emph{Tikhonov regularization.}---The Tikhonov prescription replaces $N\to N+\lambda B$ with a small positive $\lambda$, most commonly the identity matrix $B=I$~\cite{Tikhonov1977, Hansen1998}, lifting the (noisy) small eigenvalues at the cost of a regulator-dependent Hamiltonian
\begin{equation}
  H^{a}(t;\lambda) \;=\; \big(N+\lambda I\big)^{-1/2}\,H\,
                          \big(N+\lambda I\big)^{-1/2}.
  \label{eq:tik}
\end{equation}
Parameter-independent results are recovered by extrapolating $\lambda\to0$, where Eq.~\eqref{eq:tik} reduces to Eq.~\eqref{eq:Ha}. A uniform shift, however, perturbs \emph{every} eigenvalue equally, including the well-resolved large ones that need no regularization, so the extrapolation that removes it injects extra noise. To keep the modification minimal, we instead filter only the noisy subspace. Diagonalizing $N=U\Lambda U^{T}$ with eigenvalues ordered from smallest to largest, we shift only the lowest $i$ of them,
\begin{equation}
    N'(\lambda) =
U \begin{pmatrix}
\Lambda_i +\lambda I_i & 0\\
0 & \Lambda_r
\end{pmatrix}U^T ,
\label{eq:Tikhonov}
\end{equation}
where $I_i$ is the $i\times i$ identity and $r = \mathrm{dim}(N) - i$. The choice of the regularization dimension $i$ is fixed by scanning all admissible values and selecting the most stable extrapolation, we find $i=7$ (Sec.~S4~\cite{SM}). The adiabatic Hamiltonian $H^{a}(t;\lambda)=N'(\lambda)^{-1/2}\,H\,N'(\lambda)^{-1/2}$ is then extrapolated to $\lambda\to0$. The extrapolation must be carried out from an intermediate window---at very small $\lambda$ the ill-conditioning returns, while at large $\lambda$ the regulator effects dominate and suppress the physics---and the freedom in choosing that window contributes a systematic uncertainty (Sec.~S3~\cite{SM}).

\emph{Truncated SVD regularization.}---As an independent check, we replace the (soft) Tikhonov regulator with a (hard) truncated singular-value decomposition (TSVD) scheme. Again, we discard the lowest $i$ eigenvalues outright, giving
\begin{equation}
    N_{\mathrm{TSVD}} =
U \begin{pmatrix}
0_i & 0\\
0 & \Lambda_r
\end{pmatrix}U^T ,
\label{eq:TSVD}
\end{equation}
and work with the Moore-Penrose pseudo-inverse $N^{+} = U_r\Lambda_r^{-1}U_r^T$, so that $H^{a}=(N^{+})^{1/2}\, H\,(N^{+})^{1/2}$ with $(N^{+})^{1/2}=U_r\Lambda_r^{-1/2}U_r^T$. As $N^{+}$ acts within the space spanned by the radial bins, $H^{a}$ remains an $M\times M$ operator defined over those same bins. The TSVD truncation scheme discards only the near-null combinations of radial states associated with the smallest norm eigenvalues, not any radial bin itself. Truncating a fixed number of eigenvalues, rather than all those below a fixed threshold, avoids spurious noise when the threshold falls near an eigenvalue and the truncation count fluctuates across different MC samples. Scanning the admissible truncations again selects $i=7$ (Sec.~S4~\cite{SM}).

Equations~\eqref{eq:Tikhonov} and~\eqref{eq:TSVD} are closely related. As $\lambda\to\infty$ the shifted eigenvalues diverge, their contribution to the inverse vanishes, and the filtered Tikhonov construction reduces to the TSVD one. At finite $\lambda$, Tikhonov is a soft regulator and TSVD a hard truncation. The two procedures agree within statistical uncertainties (Sec.~S4~\cite{SM}), demonstrating that the ill-conditioning is confined to the small-eigenvalue subspace and that damping or removing those modes---rather than investing $\sim\!100\times$ more statistics to resolve them---recovers the physical phase shifts.

\section{Results}

\begin{figure}[t]
  \centering
  \includegraphics[width=\columnwidth]{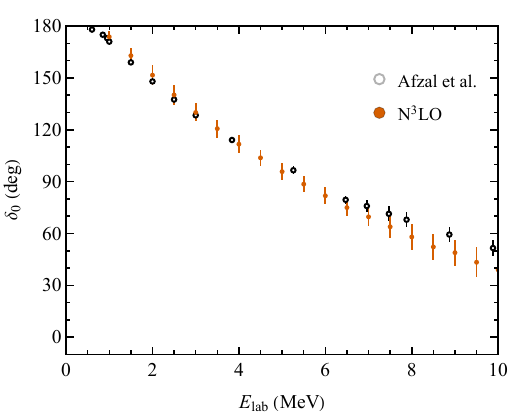}
  \caption{$S$-wave $\alpha$--$\alpha$ phase shift versus laboratory energy. Open circles: empirical phase shifts~\cite{Afzal1969}. Filled circles: current N$^{3}$LO lattice (theory) results, using Tikhonov regularization, with regulator removed by extrapolation. Theory error bars are jackknife estimates.} 
  \label{fig:swave}
\end{figure}

\begin{figure}[t]
  \centering
  \includegraphics[width=\columnwidth]{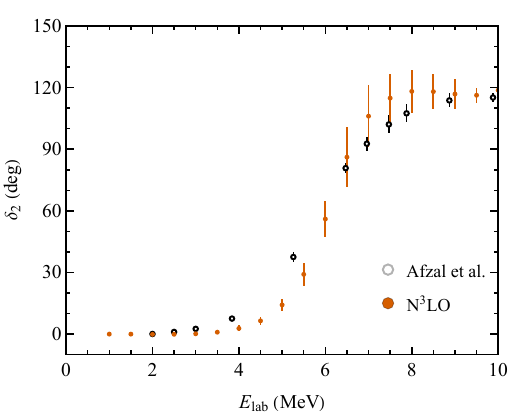}
  \caption{$D$-wave $\alpha$--$\alpha$ phase shift versus laboratory energy. Notation as in Fig.~\ref{fig:swave}. The phase shift passes through $90^{\circ}$ near the $2^{+}$ resonance of $^{8}$Be.}
  \label{fig:dwave}
\end{figure}

We perform lattice MC simulations at Euclidean-time projections $L_t = 20,\,30,\,40,\,50,$ and $60$ (lattice units), construct and regularize the adiabatic Hamiltonian, extract the phase shifts with the spherical-wall method, and extrapolate to $L_t\to\infty$ at each energy (Sec.~S5~\cite{SM}). Figures~\ref{fig:swave} and~\ref{fig:dwave} show the resulting $S$- and $D$-wave phase shifts together with the empirical analysis~\cite{Afzal1969}. The $S$-wave phase shift begins near $180^{\circ}$ at threshold---reflecting the deeply bound Pauli-forbidden configurations of the eight-nucleon system, in accordance with Levinson's theorem---and decreases smoothly with energy. The $D$-wave phase shift rises from zero through $90^{\circ}$ near the $^{8}$Be $2^{+}$ resonance. In both partial waves, the lattice results track the empirical phase shifts across the energy range studied, with no adjustable scattering parameters. An effective-range-expansion (ERE) fit~\cite{BetheERE, Elhatisari2016APM} to the $D$-wave yields the resonance parameters $E_R=3.25(25)$~MeV and $\Gamma_R=0.75(25)$~MeV, in reasonable agreement with the empirical values $E_R = 2.92(18)$~MeV and $\Gamma_R = 1.34(50)$~MeV~\cite{Afzal1969}. This represents a significant improvement over the N$^{2}$LO analysis of Ref.~\cite{Elhatisari2015AA}.

The regularization is necessary. A brute-force inversion of $N$ gives central values consistent with the data, but with uncertainties so large as to be uninformative. Crucially, the soft Tikhonov and hard truncated-SVD extractions agree within their jackknife uncertainties across the full energy range (Sec.~S4~\cite{SM}). Because the two schemes treat the small-eigenvalue subspace in opposite ways---Tikhonov damps it smoothly while TSVD removes it outright---this agreement shows that the physical phase shifts are controlled by the well-resolved, large-eigenvalue subspace of the norm matrix and are insensitive to the noisy low-norm directions. The truncated-SVD points nonetheless scatter somewhat more than the Tikhonov ones, most visibly near the resonance. The hard cut discards the small-eigenvalue subspace entirely, including whatever physical signal is admixed into it, whereas the soft damping retains a denoised estimate of that contribution through the $\lambda\to0$ extrapolation. The rank cut is also more sensitive to sample-to-sample reordering of the smallest eigenvalues. We therefore quote the Tikhonov extraction as our final result.

Two residual discrepancies are visible in Figs.~\ref{fig:swave} and~\ref{fig:dwave}. The $S$-wave falls slightly below the empirical analysis above $\sim\!7$~MeV, and the $D$-wave rises somewhat too slowly below the resonance. In both cases the calculation underpredicts the phase shift. These are the regimes most sensitive to the short-range, off-shell content of the interaction---precisely the physics that elastic scattering probes beyond bound-state observables---and we expect them to respond to the three-nucleon force, whose importance for the closely related $n$--$\alpha$ system was emphasized in Ref.~\cite{Elhatisari_2025}. For the $D$-wave, the effect is compounded at low energy by the centrifugal barrier, which suppresses the two-cluster wave function in the interaction region, so the signal there is intrinsically weak and more vulnerable to residual lattice and stochastic systematics. The large uncertainties in the $D$-wave resonance region have a separate, well-understood origin. The lattice data here approach the $L_t\to\infty$ limit only slowly, so the Euclidean-time extrapolation is poorly constrained at the projection times available (Sec.~S5~\cite{SM}).

\section{Summary and outlook}

We have presented the first \textit{ab initio} $\alpha$--$\alpha$ scattering calculation using the chiral interaction at N$^{3}$LO on a fine lattice with $a=1.32~\mathrm{fm}$. Both the main difficulty and its resolution are methodological. Reducing the eight-nucleon system to an effective two-cluster problem with the APM, we found the two-cluster norm matrix to be severely ill-conditioned at this fine lattice spacing, so that its direct inversion is unstable. Treating the inversion as an ill-posed problem and applying Tikhonov regularization, checked against an independent truncated-SVD construction, recovers precise $S$- and $D$-wave phase shifts in agreement with experiment and with no free parameters.

The agreement is good but not yet perfect, and the residual discrepancies discussed above point to a refinement of the three-nucleon interaction. Guided by the same tension in the $n$--$\alpha$ system~\cite{Elhatisari_2025}, we are revisiting the $3N$ sector, and a future $\alpha$--$\alpha$ calculation with the improved force should sharpen the comparison with the empirical phase shifts.

A complementary direction is to improve the conditioning of the norm matrix at its source, before any regularization is applied. The key observation is that damping or removing the lowest $i$ eigenvalues of $N$ is \emph{not} the same as discarding $i$ radial bins. Each norm eigenvector is a linear combination of the radial bin states (the columns of $U$), so every bin continues to contribute, only with reweighted coefficients. The ill-conditioning is therefore not localized to particular bins but reflects near-linear dependences among neighboring bins that the MC noise cannot resolve, and the small-eigenvalue eigenvectors identify which radial regions are over-resolved relative to the stochastic precision. This information can be fed back into the basis construction. Bins the noise cannot distinguish can be merged into a single wider bin, while cleanly resolved regions can be binned more finely. Such an adaptive, non-uniform binning would yield a norm matrix that is well-conditioned by construction, removing the need for \textit{a posteriori} regularization altogether. Designing the optimal binning is itself nontrivial, since the merged basis must still resolve the asymptotic region required by the spherical-wall extraction, but it offers a promising path to higher-precision APM calculations at fixed MC cost.

More broadly, our results extend the validation of this high-fidelity chiral interaction~\cite{Elhatisari2024WFM, Shen2025Be} from bound-state energies and charge radii~\cite{Ren:2025vpe} to scattering processes, and show that high-fidelity chiral forces on a fine lattice describe $\alpha$-cluster dynamics quantitatively. The regularization strategy removes an important practical obstacle for the APM at fine lattice spacing, opening the way to further \textit{ab initio} calculations of reactions of nuclear and astrophysical relevance. Work applying these methods to deuteron--deuteron scattering~\cite{helen} and radiative $\alpha$ capture is in progress.

\begin{acknowledgments}
We thank the members of the NLEFT collaboration for valuable discussions. This work was supported in part by the European Research Council (ERC) under the European Union's Horizon 2020 research and innovation programme (grant agreement No. 101018170) and by the CAS President's International Fellowship Initiative (PIFI) (Grant No.~2025PD0022). The authors gratefully acknowledge the Gauss Centre for Supercomputing e.V. (www.gauss-centre.eu) for funding this project by providing computing time on the GCS Supercomputers JUWELS and JUPITER at J\"ulich Supercomputing Centre (JSC) and the support of the project \mbox{EXOTIC} by the JSC through dedicated HPC time on the \mbox{JURECA DC} GPU partition.
Furthermore, the authors gratefully acknowledge the computing time provided on the high-performance computer
HoreKa by the National High-Performance Computing Center at KIT (NHR@KIT). This center is
jointly supported by the Federal Ministry of Education and Research and the Ministry of Science,
Research and the Arts of Baden-W\"urttemberg, as part of the National High-Performance Computing
(NHR) joint funding program (https://www.nhr-verein.de/en/our-partners). HoreKa is partly funded
by the German Research Foundation (DFG).
\end{acknowledgments}

\bibliography{refs}

\clearpage
\onecolumngrid
\begin{center}
  \textbf{\large Supplemental Material:\\
  Ab initio $\alpha$--$\alpha$ scattering with high-fidelity chiral interactions}
\end{center}
\setcounter{equation}{0}
\setcounter{figure}{0}
\setcounter{table}{0}
\renewcommand{\theequation}{S\arabic{equation}}
\renewcommand{\thefigure}{S\arabic{figure}}
\renewcommand{\thetable}{S\arabic{table}}

This Supplemental Material collects the technical details underlying the Letter: the interaction and lattice setup (Sec.~S1), the construction of the two-cluster basis and the two-step adiabatic Hamiltonian (Sec.~S2), the Tikhonov $\lambda\to0$ extrapolation (Sec.~S3), the choice of the regularization dimension and the Tikhonov/TSVD comparison (Sec.~S4), and the Euclidean-time extrapolation (Sec.~S5).

\subsection*{S1. Hamiltonian and lattice conventions}
The calculations presented here are based on the ``high-fidelity'' chiral interaction introduced in Ref.~\cite{Elhatisari2024WFM}. It should
be noted that our recent $n$--$\alpha$ companion study of Ref.~\cite{Elhatisari_2025} uses the same interaction as the current work. Our interaction can be expressed as
\begin{equation}
H = K + V_\mathrm{OPE}^{}
+ V_\mathrm{C}^{} + V_\mathrm{2N}^{Q^4}
+ V_\mathrm{3N}^{Q^3}
+ W_\mathrm{2N}^{Q^4},
\end{equation}
where $K$ is the kinetic energy term and $V_\mathrm{OPE}$ the one-pion exchange potential. The long-range Coulomb repulsion between protons is represented by $V_\mathrm{C}$ and included throughout. The two-nucleon force $V_\mathrm{2N}^{Q^4}$
is included through N$^{3}$LO, with the two-pion-exchange contributions absorbed into four-nucleon contact terms. This is
supplemented by a smeared N$^{2}$LO three-nucleon force $V_\mathrm{3N}^{Q^3}$. Furthermore, $W_\mathrm{2N}^{Q^4}$ encodes 
the $2N$ Galilean invariance restoration interaction at N$^{3}$LO.

The N$^{3}$LO high-fidelity interaction is incorporated using wave function matching~\cite{Elhatisari2024WFM}, a short-range unitary transformation, $H^{\prime} = U^{\dagger} \, H\, U$ that maps normalized orthogonal states of $H$ to normalized orthogonal states of a computationally simple Hamiltonian $H_S$, which results in perturbation friendly residual corrections $H^{\prime}-H_S$. It should be noted that $H_S$ is largely free of
the MC sign problem. Our high-fidelity action reproduces a broad set of observables, including $NN$ phase shifts, the binding energies and charge radii of nuclei up to $A\leq 58$, the neutron-matter equation of state, and the saturation properties of nuclear matter~\cite{Elhatisari2024WFM}. The full specification of the action, the low-energy constants, and the MC updating scheme are given in Refs.~\cite{Elhatisari2024WFM, Elhatisari_2025}.

The calculations are performed on a spatial lattice with spacing $a=1.32~\mathrm{fm}$, corresponding to a momentum cutoff $\Lambda=\pi/a\simeq 470~\mathrm{MeV}$, and temporal lattice spacing $a_t=(1000~\mathrm{MeV})^{-1}$. The fully interacting eight-nucleon adiabatic Hamiltonian is computed on a periodic cubic box of side length $L'=10$ lattice units. Separately, the asymptotic two-cluster adiabatic Hamiltonian, in which the two $\alpha$ clusters are non-interacting except for the long-range Coulomb interaction, is computed on a much larger periodic cubic box of side length $L=72$ lattice units. The two Hamiltonians are then matched in the asymptotic region, where the short-range nuclear interaction between the clusters has vanished, to obtain the large-volume adiabatic Hamiltonian used in the spherical-wall analysis (see Sec.~S2). The cluster states are projected over Euclidean times $L_t a_t=0.02$--$0.06~\mathrm{MeV}^{-1}$, corresponding to $L_t=20$--$60$ in lattice units, and the resulting phase shifts are extrapolated to $L_t\to\infty$ as described in Sec.~S5.

\subsection*{S2. Two-cluster basis and the two-step adiabatic Hamiltonian }
We summarize the APM and the associated techniques used to extract scattering information. All of this follows Ref.~\cite{Elhatisari2016APM}, and the implementation here is identical, but we collect the key steps for the reader's convenience.

For generality, consider two clusters of $A_{1}$ and $A_{2}$ nucleons. The APM starts from initial cluster states defined on the lattice and evolves them in Euclidean projection time. One starts with an approximate two-cluster description and lets the microscopic interaction dress it into the physical low-lying cluster states, including all the deformation and polarization the interaction induces. We define the initial Slater-determinant cluster states as
\begin{equation}
\ket{\vec{d}\,\,} = \sum_{\vec{n}}
\ket{\vec{n}+\vec{d}\,\,}_{1} \otimes \ket{\vec{n}}_{2} \,,
\label{eqn:InitialClusterStates-001}
\end{equation}
characterized by the two-cluster displacement $\vec{d}$. We project these onto spherical harmonics ${\rm{Y}}_{\ell,\ell_{z}}$,
\begin{equation}
 \ket{d}^{\ell,\ell_{z}} =
 \sum_{\vec{d}^{\prime}}
 {\rm{Y}}_{\ell,\ell_{z}}(\hat{d}\,')
 \delta_{d,|\vec{d}\,'|}
 \ket{\vec{d}\,'} \,,
 \label{eqn:InitialClusterStates-005}
\end{equation}
which amounts to defining the cluster states in radial coordinates. We then bin the lattice points into radial (annular) shells of width $a_{R}$, grouping points equidistant in $|\vec{d}|$, and let $R$ denote the midpoint of each bin, so that
\begin{equation}
 \ket{R}^{\ell,\ell_{z}} =
 \sum_{|d-R|<{a_{R}}/{2}}
 \ket{d}^{\ell,\ell_{z}}\,.
 \label{eqn:InitialClusterStates-009}
\end{equation}
This binning is what makes large-scale calculations tractable. Table~I of Ref.~\cite{Elhatisari2016APM} shows, for example, that for $L=10$ a $1000\times1000$ two-cluster adiabatic matrix constructed from the cubic-lattice cluster states in Eq.~(\ref{eqn:InitialClusterStates-001}) is reduced to $14\times14$ when the cluster states in Eq.~(\ref{eqn:InitialClusterStates-009}) are used.

The initial cluster states satisfy the completeness relation
\begin{equation}
I = \sum_{{R},{R}^{\prime}}
\ket{R}^{\ell,\ell_{z}}
\
[N_{0}^{-1}]_{{R},{R}^{\prime}}^{\ell,\ell_{z}}
\
\prescript{\ell,\ell_{z}}{}{\bra{R^{\prime}}}
\,,
\label{eqn:CompletenesRel-001}
\end{equation}
with the initial norm matrix
\begin{equation}
[N_{0}]_{R,R^{\prime}}^{\ell,\ell_{z}} =
\prescript{\ell,\ell_{z}}{}{\braket{R|R^{\prime}}}^{\ell,\ell_{z}}\,,
\label{eqn:NormMatrix-001}
\end{equation}
whose elements measure the overlap between cluster states separated by $|\vec{R}|$ and $|\vec{R}^{\prime}|$. Using the transfer-matrix operator $\hat{M}=:\exp(-\hat{H}\,a_t):$, we evolve the initial states in Euclidean time into the dressed states
\begin{equation}
\ket{R}^{\ell,\ell_{z}}_{{n_{t}}}
=
\hat{M}^{n_{t}}
\,
\ket{R}^{\ell,\ell_{z}}\,,
\label{eqn:InitialClusterStates-013}
\end{equation}
with $n_{t} = L_{t}/2$, and form norm and Hamiltonian matrices at a finite Euclidean time,
\begin{equation}
[N_{L_{t}}]_{R,R^{\prime}}^{\ell,\ell_{z}} =  \prescript{\ell,\ell_{z}}{n_{t}}{\braket{R|R^{\prime}}}_{n_{t}}^{\ell,\ell_{z}},
\label{eqn:SM_NormMatrix}
\end{equation}
\begin{equation}
[H_{L_{t}}]_{R,R^{\prime}}^{\ell,\ell_{z}} =  \prescript{\ell,\ell_{z}}{n_{t}}{\braket{R|\hat{H}|R^{\prime}}}_{n_{t}}^{\ell,\ell_{z}}.
\label{eqn:SM_Hamiltonian}
\end{equation}
Because the dressed states are not orthogonal, the adiabatic Hamiltonian is obtained from the norm matrix as
\begin{equation}
[H_{L_{t}}^{\text{a}}]^{\ell,\ell_{z}}_{R,R^{\prime}} =
[N_{L_{t}}^{-1/2}\,H_{L_{t}}\,N_{L_{t}}^{-1/2}]_{R,R^{\prime}}^{\ell,\ell_{z}},
\label{eqn:SM_AdiabaticHamiltonian}
\end{equation}
and encodes all the low-energy physics of the two-cluster system. Phase shifts are then extracted from the adiabatic Hamiltonian with the spherical-wall method~\cite{Borasoy2007, Lu2016, Elhatisari2016APM}.

\begin{figure}[htbp]
    \centering
    \begin{subfigure}[b]{\textwidth}
        \centering
        \includegraphics[width=\textwidth]{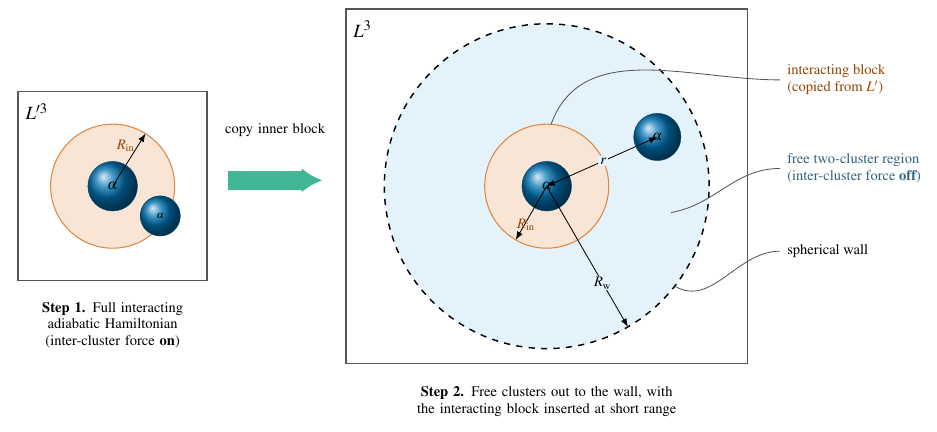}
        \caption{Real-space picture}
        \label{fig:panel1}
    \end{subfigure}
    \hfill
    \begin{subfigure}[b]{0.8\textwidth}
        \centering
        \includegraphics[width=\textwidth]{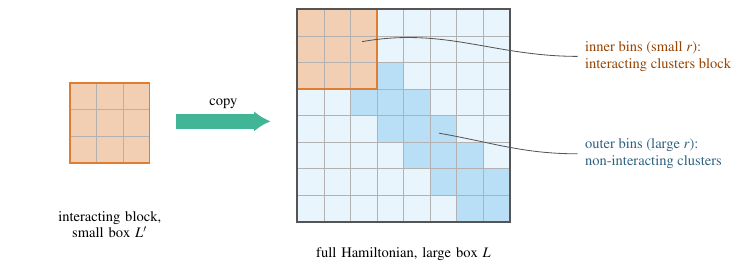}
        \caption{Matrix picture}
        \label{fig:panel2}
    \end{subfigure}
    \caption{Two-step construction of the adiabatic Hamiltonian. The full interacting adiabatic Hamiltonian (computed on a small lattice) is inserted into the non-interacting adiabatic Hamiltonian (computed on a large lattice with the inter-cluster interaction switched off) to obtain the large-lattice adiabatic Hamiltonian from which the phase shifts are read off. Note that we impose Coulomb boundary conditions on the spherical wall to account for the long-range Coulomb interaction between the two $\alpha$-particles.}
    \label{fig:SM_Hamiltonian_matching}
\end{figure}

Extracting low-energy phase shifts requires $H^{a}$ on a large lattice, but the full interacting eight-nucleon calculation is feasible only on a small one. The short range of the interaction lets us split the construction into two (Fig.~\ref{fig:SM_Hamiltonian_matching}). First, the full interacting adiabatic Hamiltonian is computed on a periodic lattice of volume $L'^{3}\times L_t$, larger than the interaction range (but not so large as to be computationally unfeasible). Second, the adiabatic Hamiltonian of the individual clusters---with all intra-cluster interactions on but the inter-cluster interaction off---is computed on a large lattice of volume $L^{3}\times L_t$ with $L\gg L'$. This non-interacting adiabatic Hamiltonian is exact in the asymptotic region. We then insert the interacting block into the non-interacting Hamiltonian to obtain the large-lattice $H^{a}$.

The interacting Hamiltonian block must merge smoothly into the non-interacting Hamiltonian. We test this by matching successively larger interacting blocks $(L'-n)^{3}\times L_t,\,(L'-n+1)^{3}\times L_t,\,\dots, \, L'^{3}\times L_t$: a perfect match gives the same phase shift for each. In practice, the Coulomb interaction makes the interacting and non-interacting Hamiltonians differ, especially at small block sizes, so we add a small diagonal offset to the interacting adiabatic Hamiltonian and tune it so that the different matchings coincide. This convergence with respect to the matching size is an important diagnostic of a properly constructed $H^{a}$, and it plays a central role once the (Tikhonov or TSVD) regularization is introduced in the following sections.

We use $11$ radial bins on the interacting box, but the innermost bins have negligible support and must be merged before the norm matrix can be inverted. At small separation, the two $\alpha$ clusters overlap, and the antisymmetrization of the eight nucleons strongly suppresses the corresponding cluster state, so its norm nearly vanishes. For the $D$-wave, the $\ell=2$ angular projection suppresses small separations still further. A norm matrix built on the full $11$-bin basis, therefore, has near-null rows and columns and is numerically singular for reasons that are entirely deterministic and unrelated to MC statistics. We remove this trivial singularity by merging the empty inner bins---the first bin into the second for the $S$-wave, and the first three into a single bin for the $D$-wave---which leaves an effective interacting Hamiltonian of dimension $10$ and $9$, respectively. The same merge is applied to the large-box non-interacting Hamiltonian, so that the interacting and non-interacting Hamiltonians share a common radial basis when they are matched, its dimension is reduced correspondingly from $72$ to $71$ for the $S$-wave and to $70$ for the $D$-wave. This deterministic reduction should not be confused with the \emph{stochastic} ill-conditioning of the remaining basis, which is the subject of Secs.~S3 and~S4.

\subsection*{S3. Tikhonov $\lambda\to0$ extrapolation}
As discussed in the main text, on the fine lattice, the dressed cluster states become nearly linearly dependent as they are projected in Euclidean time, and $N$ becomes ill-conditioned, so a direct inversion amplifies the MC noise, and the brute-force route would require more than $100$ times the present statistics. We regularize the inversion with a Tikhonov parameter $\lambda$,
\begin{equation}
  H^{a}(t;\lambda) \;=\; \big(N+\lambda I\big)^{-1/2}\,H\,\big(N+\lambda I\big)^{-1/2},
  \label{eq:SM_tik}
\end{equation}
with the physical result being the limit $\lambda\to0$. As in the main text, we apply the shift only to the lowest $i$ eigenvalues,
\begin{equation}
    N'(\lambda) =
U \begin{pmatrix}
\Lambda_i +\lambda I_i & 0\\
0 & \Lambda_r
\end{pmatrix}U^T .
\label{eq:SM_Tikhonov}
\end{equation}
The choice of $i$ is discussed in Sec.~S4. Here, we focus on the $\lambda\to0$ extrapolation.

\begin{figure}[t]
  \centering
  \includegraphics[width=0.6\columnwidth]{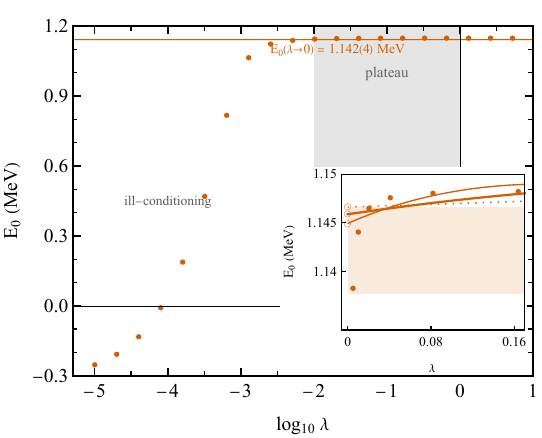}
  \caption{Lowest eigenvalue $E_{0}$ of the adiabatic Hamiltonian in the spherical-wall box, measured relative to the $\alpha$--$\alpha$ threshold, as a function of the Tikhonov parameter $\lambda$ (logarithmic axis). For $\log_{10}\lambda \lesssim -2$ the ill-conditioned inversion amplifies stochastic noise and $E_{0}$ is unstable, even turning negative---unphysical, since $^{8}$Be is unbound and has no two-$\alpha$ bound state below threshold, a broad plateau with $E_{0}>0$ develops over $-2 \lesssim \log_{10}\lambda \lesssim 0$. The $\lambda\to0$ limit is obtained by fitting $E_{0}(\lambda)$ with a low-order polynomial across the plateau and evaluating the intercept (inset, linear $\lambda$ axis; dotted, solid, and dashed curves are linear, quadratic, and cubic fits, with open symbols at $\lambda=0$). The shaded band is the conservative envelope from varying the polynomial order and the fit window, giving $E_{0}(\lambda \to 0) = 1.146(4)~\mathrm{MeV}$.}
  \label{fig:Tikhonov_extrapolation}
\end{figure}

With the spherical-wall method, the phase shifts follow directly from the eigenvalues of $H^{a}$, so the $\lambda$ dependence of the phase shifts mirrors that of the adiabatic eigenvalues. Figure~\ref{fig:Tikhonov_extrapolation} shows the lowest eigenvalue $E_0$ of a representative $D$-wave adiabatic Hamiltonian, measured relative to the $\alpha$--$\alpha$ threshold, as a function of $\lambda$. The $S$-wave behaves similarly, as do other projection times. For $\log_{10}\lambda \lesssim -2$ the ill-conditioning re-emerges and $E_0$ becomes erratic, even turning negative. Since $^{8}$Be is unbound, a negative relative energy is unphysical---it would correspond to a two-$\alpha$ state bound below threshold---so this sign change is itself a clear signature of the breakdown. For $\log_{10}\lambda \gtrsim 0$ over-regularization biases the spectrum. Between these regimes, $E_0(\lambda)$ exhibits a broad plateau, and we take the $\lambda\to0$ value by fitting a low-order polynomial in $\lambda$ across the plateau and evaluating the intercept at $\lambda=0$. Two choices contribute to systematic uncertainty: the polynomial order and the fit window. In the example shown, varying the order among linear, quadratic, and cubic shifts the intercept by less than $2~\mathrm{keV}$, whereas varying the window between $-2 \le \log_{10}\lambda \le 0$ and $-2.5 \le \log_{10}\lambda \le -0.5$ shifts it by roughly $7~\mathrm{keV}$. The latter dominates because the wider window reaches into the onset of the ill-conditioned region. Assigning a conservative systematic equal to the full envelope of intercepts gives $E_{0}(\lambda \to 0) = 1.146(4)~\mathrm{MeV}$.

\subsection*{S4. Regularization dimension and Tikhonov/TSVD comparison }
\begin{figure}[t]
  \centering
  \includegraphics[width=\columnwidth]{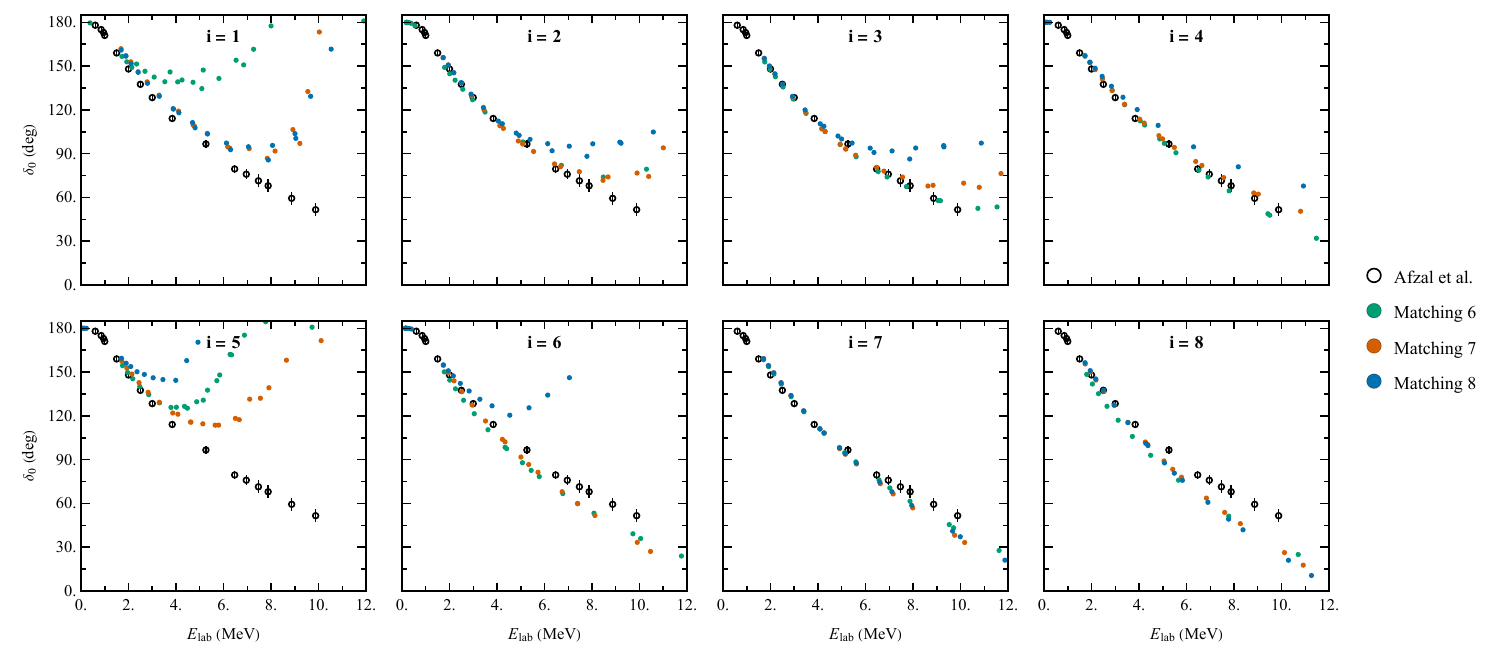}
  \caption{$S$-wave $\alpha$--$\alpha$ phase shift $\delta_{0}$ versus laboratory energy $E_{\mathrm{lab}}$ for eight choices of the Tikhonov regularization dimension (panels $i=1$--$8$, where $\lambda$ times an $i$-dimensional identity is added to the smallest norm eigenvalues). Within each panel, the colored points embed an $m\times m$ block of the fully interacting adiabatic Hamiltonian ($10\times10$) into the single-cluster adiabatic Hamiltonian ($71\times71$), for matching dimensions $m=6$--$8$. Open circles are the empirical phase shifts~\cite{Afzal1969}. A good match shows convergence in $m$, i.e.\ the $m=6$--$8$ curves collapse. The choice $i=7$ achieves the best convergence, while the others show a clear spread.}
  \label{fig:Tikhonov_choice_swave}
\end{figure}

\begin{figure}[t]
  \centering
  \includegraphics[width=\columnwidth]{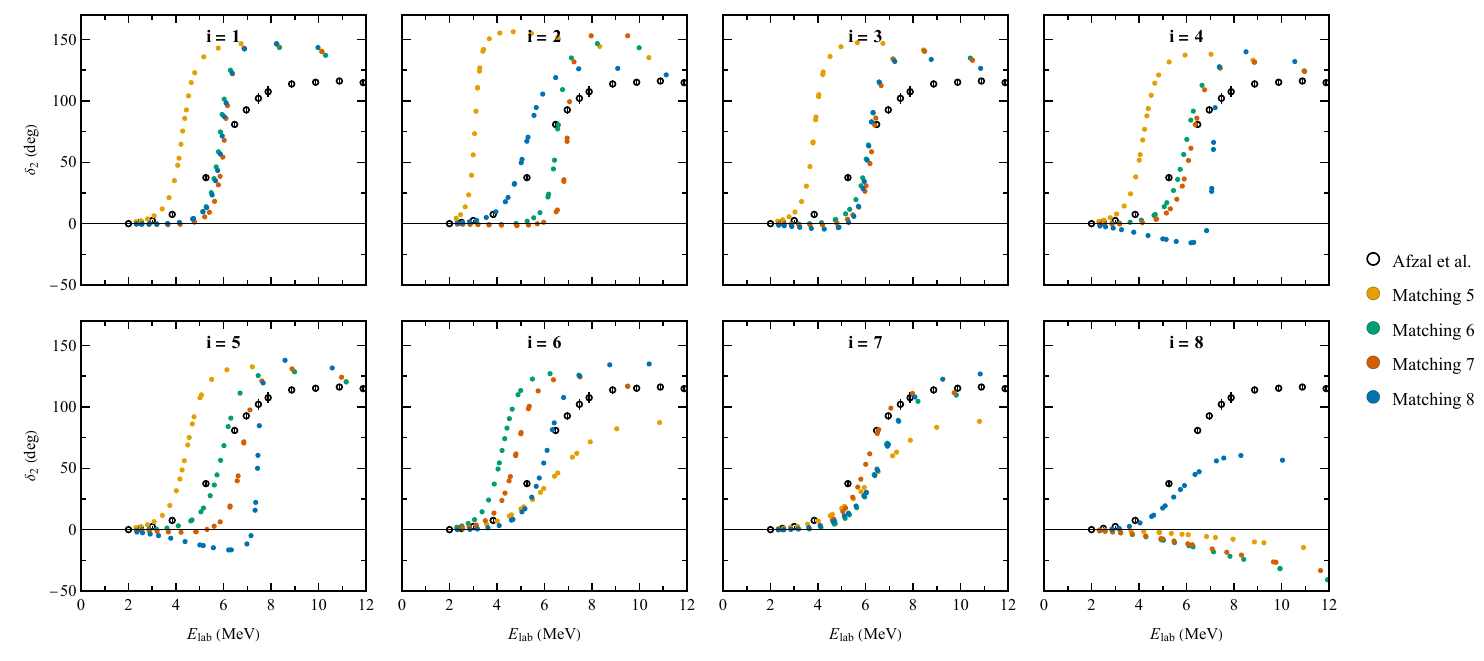}
  \caption{$D$-wave phase shift $\delta_{2}$ versus $E_{\mathrm{lab}}$ for eight choices of the Tikhonov regularization dimension. As in Fig.~\ref{fig:Tikhonov_choice_swave}, each panel embeds an $m\times m$ block of the fully interacting adiabatic Hamiltonian (here $9\times9$) into the single-cluster adiabatic Hamiltonian (here $70\times70$) for $m=5$--$8$, with open circles the empirical phase shifts~\cite{Afzal1969}. Phase shifts are defined modulo $180^{\circ}$ and drawn on the branch that varies smoothly with $E_{\mathrm{lab}}$. Points from poorly conditioned fits (uncertainty $\gtrsim 20^{\circ}$) are omitted. Again, $i=7$ gives the best convergence across matchings.}
  \label{fig:Tikhonov_choice_dwave}
\end{figure}

To fix $i$ in Eq.~\eqref{eq:SM_Tikhonov}, we compute the phase shifts for all allowed values. After the inner-bin merging as described in Sec.~S2, the interacting Hamiltonian is $10$-dimensional for the $S$-wave case and $9$-dimensional for the $D$-wave case. Figs.~\ref{fig:Tikhonov_choice_swave} and~\ref{fig:Tikhonov_choice_dwave} show the $\lambda\to0$ phase shifts for $1\leq i\leq 8$. Error bars are suppressed for clarity, and the diagonal matching offset (Sec.~S2) is optimized for the best convergence across matching sizes.

For Tikhonov regularization, the choice $i=7$ brings the different matchings closest together, indicating the most consistent matching of the interacting and non-interacting adiabatic Hamiltonians.  The final $S$-wave result is averaged over matchings $m=6,7,8$. For the $D$-wave, a residual spread remains, but the errors are smallest for $m=7$, and this matching is also the most stable across both Tikhonov and TSVD, so for the final $D$-wave result, we adopt $i=7,\,m=7$.

\begin{figure}[t]
  \centering
  \includegraphics[width=0.85\columnwidth]{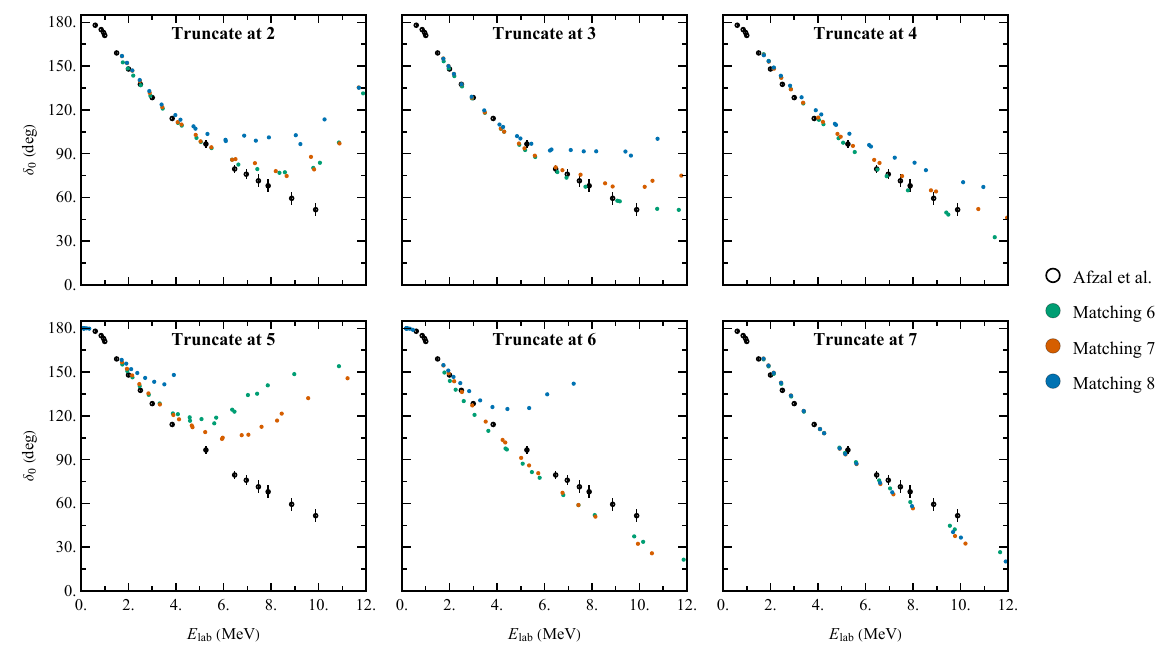}
  \caption{$S$-wave phase shift $\delta_{0}$ versus $E_{\mathrm{lab}}$ for six choices of the TSVD truncation dimension $i$.}
  \label{fig:TSVD_choice_swave}
\end{figure}

\begin{figure}[t]
  \centering
  \includegraphics[width=0.85\columnwidth]{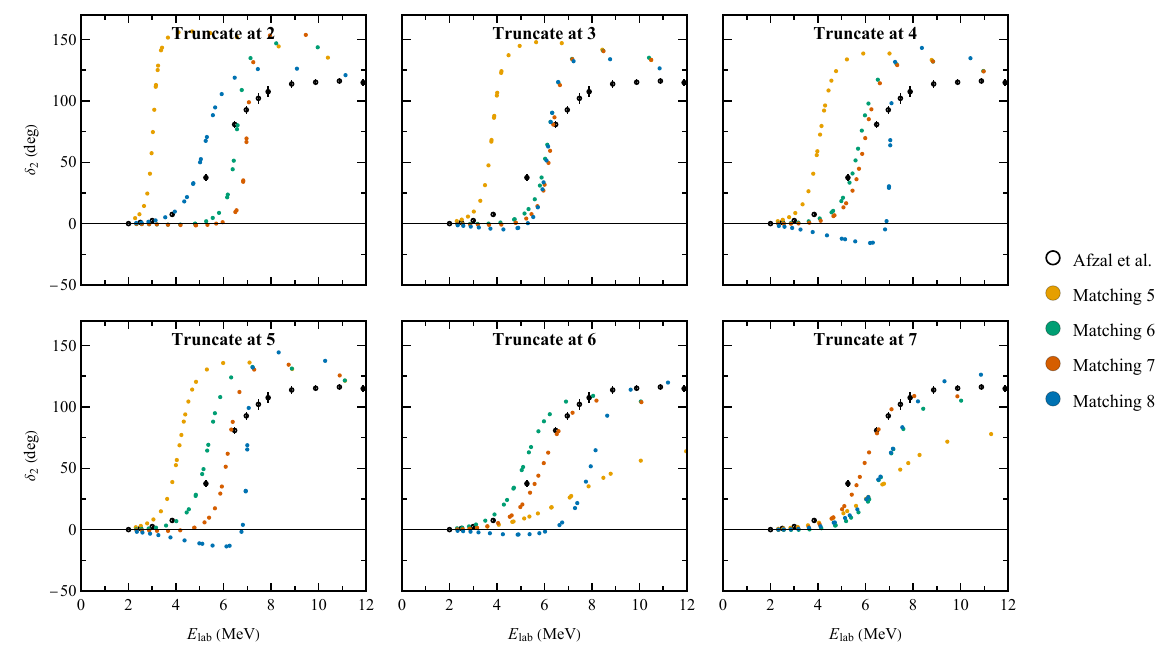}
  \caption{$D$-wave phase shift $\delta_{2}$ versus $E_{\mathrm{lab}}$ for six choices of the TSVD truncation dimension $i$.}
  \label{fig:TSVD_choice_dwave}
\end{figure}

As an independent benchmark and a test for over-regularization, we reproduce the phase shifts with the truncated-SVD construction of the main text,
\begin{equation}
    N_{\mathrm{TSVD}} =
U \begin{pmatrix}
0_i & 0\\
0 & \Lambda_r
\end{pmatrix}U^T ,
\label{eq:SM_TSVD}
\end{equation}
with the inverse defined through the Moore-Penrose pseudo-inverse $N^{+} = U_r\Lambda_r^{-1}U_r^T$, so that $H^{a}=(N^{+})^{1/2}\, H\,(N^{+})^{1/2}$ with $(N^{+})^{1/2} = U_r\Lambda_r^{-1/2}U_r^T$. We again determine $i$ by considering all truncations. Figs.~\ref{fig:TSVD_choice_swave} and~\ref{fig:TSVD_choice_dwave} show a comparison between $2\leq i\leq 7$ for the $S$- and $D$-wave. We omit the cases of $i=1$ and $i=8$ as these give poor results. As for the Tikhonov regularization, $i=7$ gives the best match in both partial waves.

We stress that removing $7$ norm eigenvalues is not the same as removing $7$ radial channels. Each norm eigenstate is a linear combination of the radial channels (given by the columns of $U$), so all channels still contribute to the final result. The ill-conditioning (from the nearly linearly dependent cluster states and the MC noise) simply leaves only a few cleanly resolvable directions, which the regularization isolates. This observation also motivates the adaptive non-uniform binning discussed in the main text.

\begin{figure}[t]
  \centering
  \includegraphics[width=0.6\columnwidth]{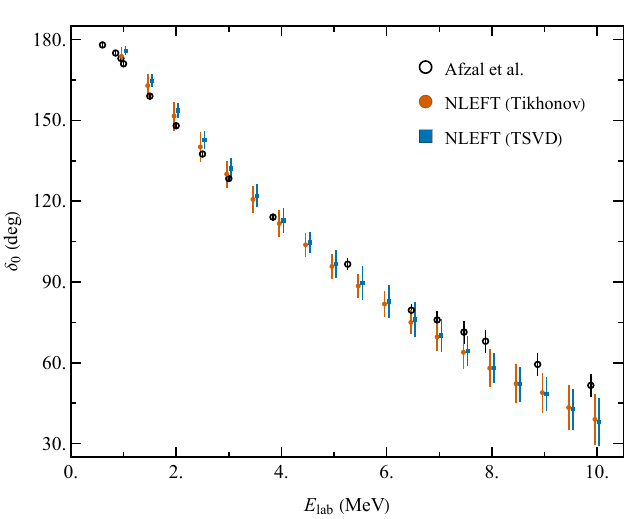}
  \caption{$S$-wave $\alpha$--$\alpha$ phase shift from the Tikhonov (filled circles) and truncated-SVD (filled squares) regularizations, compared with the empirical analysis~\cite{Afzal1969} (open circles). The two agree within jackknife uncertainties. The lattice series are offset slightly in energy for visibility.}
  \label{fig:TSVD_vs_TikhonovS}
\end{figure}

\begin{figure}[t]
  \centering
  \includegraphics[width=0.6\columnwidth]{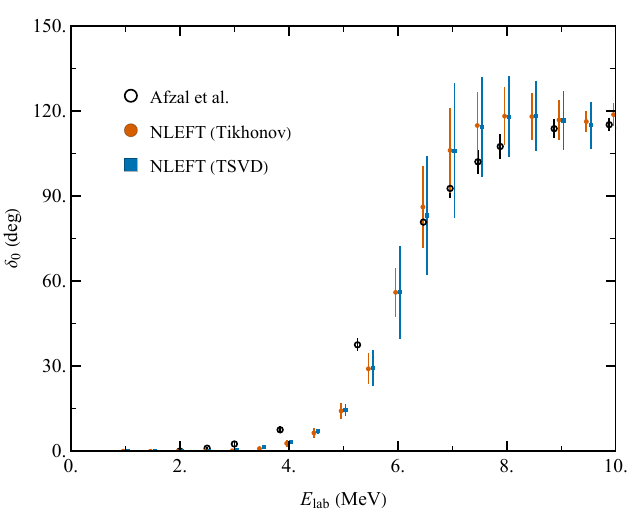}
  \caption{As in Fig.~\ref{fig:TSVD_vs_TikhonovS} for the $D$-wave.}
  \label{fig:TSVD_vs_TikhonovD}
\end{figure}

Figs.~\ref{fig:TSVD_vs_TikhonovS} and~\ref{fig:TSVD_vs_TikhonovD} compare the $S$- and $D$-wave phase shifts from the soft Tikhonov regularization [Eq.~\eqref{eq:SM_Tikhonov}] and the hard truncated-SVD construction [Eq.~\eqref{eq:SM_TSVD}], together with the empirical values~\cite{Afzal1969}. The two regularizations agree within their uncertainties across the full energy range, including the resonance region where the uncertainties are largest. Because the two schemes treat the small-eigenvalue subspace very differently (Tikhonov damps it smoothly, TSVD removes it outright) this agreement shows that the physical phase shifts are controlled by the well-resolved, large-eigenvalue subspace and are insensitive to the (noisy) low-norm directions.  We present the Tikhonov phase shifts as our final result because they are more stable near the resonance: the soft regulator retains, through the $\lambda\to0$ extrapolation, a denoised estimate of the small-eigenvalue contribution that the TSVD discards entirely.

\subsection*{S5. Euclidean-time extrapolation}
\begin{figure}[t]
  \centering
  \includegraphics[width=\columnwidth]{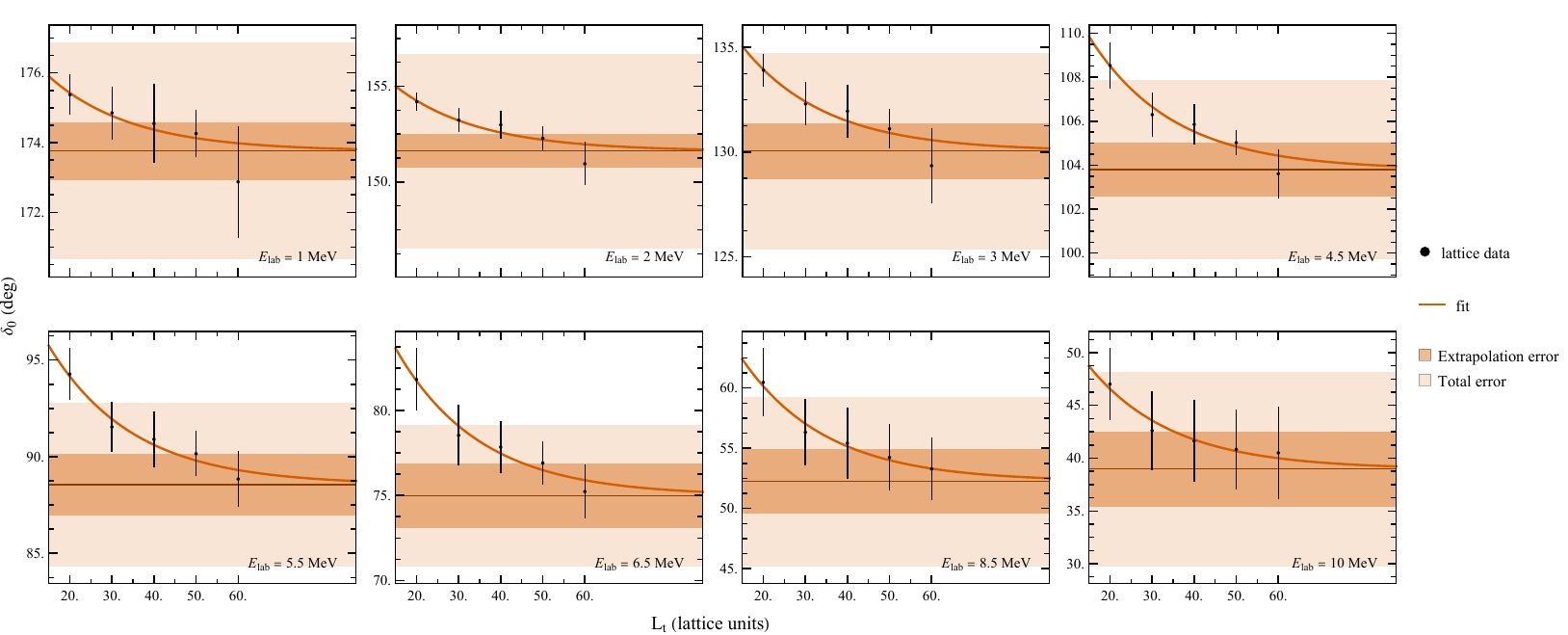}
  \caption{$L_t \to \infty$ extrapolation of the $S$-wave phase shift $\delta_{0}$ at eight laboratory energies. Points are the lattice results at $L_t = 20$--$60$, error bars are the $1\sigma$ MC uncertainty. The   solid curve is the fit $\delta_{0}(L_t) = \delta_\infty + a\,e^{-d_E L_t}$ and the dashed line marks $\delta_\infty$. The inner (darker) band is the extrapolation error, and the outer (lighter) band is the total uncertainty including systematics, the latter coinciding with the error quoted in the final result.}
  \label{fig:Swave_LtExtrapolation}
\end{figure}

\begin{figure}[t]
  \centering
  \includegraphics[width=\columnwidth]{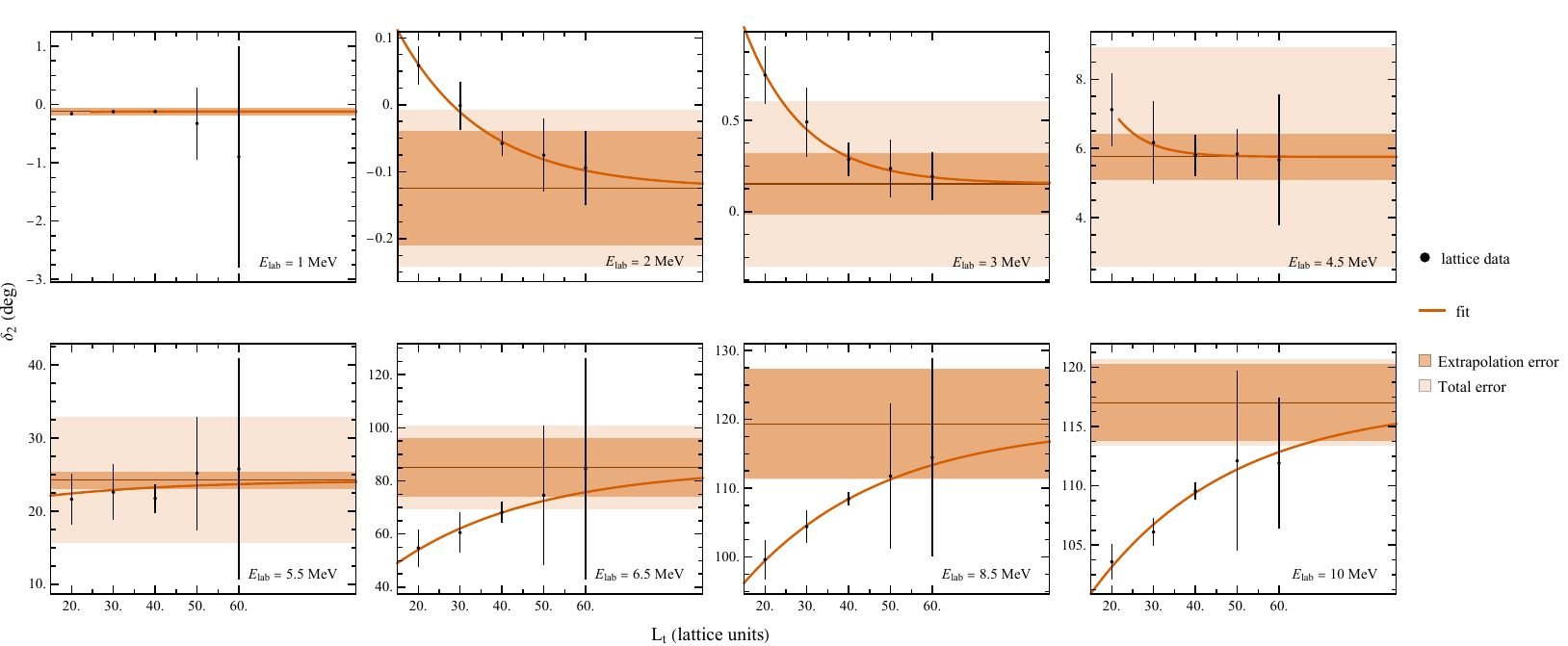}
  \caption{As in Fig.~\ref{fig:Swave_LtExtrapolation}, for the $D$-wave phase shift $\delta_{2}$, fitting $\delta_{2}(L_t)=\delta_\infty+a\,e^{-d_E L_t}$.}
  \label{fig:Dwave_LtExtrapolation}
\end{figure}

We compute the adiabatic Hamiltonian for the $S$- and $D$-wave channels at Euclidean-time projections $L_t = 20,\,30,\,40,\,50,$ and $60$ (lattice units). After calculation of the phase shifts, these are extrapolated to $L_t\to\infty$ at each energy. Typical Euclidean time extrapolations are shown in Figs.~\ref{fig:Swave_LtExtrapolation} and~\ref{fig:Dwave_LtExtrapolation}. The uncertainty at each lattice point is estimated by jackknife resampling, and we assume an exponential approach to the large Euclidean time limit (see Appendix~A of Ref.~\cite{Elhatisari2016APM}), with the extrapolation error (darker band) determined by bootstrap.  In addition, systematic uncertainties arise from the Tikhonov $\lambda\to0$ extrapolation and from the choice of the matching offset, the lighter band shows the sum of the extrapolation and systematic uncertainties, which is the error quoted in the main results.

The $D$-wave extrapolation figure shows why the resonance-region error bars are so large. At $E_{\mathrm{lab}} = 6.5$ and $8.5$~MeV, the phase shifts are still nearly linear in $L_t$ at the available projection times, so the approach to the $L_t\to\infty$ limit has not yet set in and the extrapolation is poorly constrained. Pinning it down would require data at larger $L_t$, but that is precisely where the ill-conditioning is most severe: as $L_t$ grows, the Euclidean projection drives the dressed cluster states toward the few lowest-lying two-cluster states, so neighboring radial bins become increasingly linearly dependent and the norm matrix ever closer to singular, while the MC noise on its entries grows as well. Larger projection times are therefore not accessible at the present statistics. Reaching them would require first controlling the ill-conditioning---through higher statistics or the adaptive radial binning discussed in the main text.

\end{document}